\documentclass[12pt]{article}
\pdfoutput=1

\usepackage{amsfonts}
\usepackage{amssymb}
\usepackage{graphicx}
\usepackage{latexsym}
\usepackage{amsmath}
\newcommand\openone{\leavevmode\hbox{\small1\normalsize\kern-.33em1}}
\def\ee{\end{eqnarray}}

\def\nn{\nonumber}

\newcommand{\half}{{\textstyle{\frac12}}}

\newcommand{\be}{\begin{eqnarray}}
\newcommand{\en}{\end{eqnarray}}
\newcommand{\bea}[1]{\left(\begin{array}{#1}}
\newcommand{\ena}{\end{array}\right)}
\newcommand{\CO}{\mathcal{O}}

\newcommand{\tmop}[1]{\ensuremath{\operatorname{#1}}}

\begin{document}

\vspace{5mm}
\vspace{0.5cm}
\begin{center}

\def\thefootnote{\fnsymbol{footnote}}

{\Large \bf Discovering Asymmetric Dark Matter with Anti-Neutrinos}
\\[0.5cm]
{\large Brian Feldstein,  A.~Liam Fitzpatrick}

{\small \textit{
Physics Department \\
Boston University, Boston, MA 02215, USA}}

\end{center}

\vspace{.8cm}

\hrule \vspace{0.3cm}
{\small  \noindent \textbf{Abstract} \\[0.3cm]
\noindent

We discuss possible signatures of Asymmetric Dark Matter (ADM)
through dark matter decays to neutrinos.
We specifically focus on scenarios in which
the Standard Model (SM) baryon asymmetry is transferred to the dark sector
(DS) through higher dimensional operators in chemical equilibrium.
In such cases, the dark matter (DM) carries lepton and/or baryon number, and
we point out that for a wide range of quantum number assignments,
by far the strongest constraints on dark matter decays come from
decays to neutrinos through the ``neutrino portal'' operator
$HL$.  Together with the facts that ADM favors
lighter DM masses $\sim$ a few GeV and that the decays would lead only to
anti-neutrinos and no neutrinos (or vice versa), the detection
of such decays at neutrino telescopes would provide compelling
evidence for ADM.  We discuss current and future bounds on
models where the DM decays to neutrinos through operators
of dimension $\le 6$.  For dimension 6 operators, the scale suppressing
the decay is bounded to be $\gtrsim 10^{12} - 10^{13}$ GeV.

\vspace{0.5cm}  \hrule
\def\thefootnote{\arabic{footnote}}}
\setcounter{footnote}{0}

\section{Introduction}
\label{sec:intro}

It is by now well established that about a quarter of the mass of
the universe is in the form of some kind of non-luminous, non-baryonic matter.
%
%
The standard cosmic history leading to thermal relic dark matter (DM),
where DM annihilations freeze out at temperatures near its mass,
is certainly an attractive possibility, but it is not the only one.
Indeed, the observed baryon and lepton number densities did not arise due to
thermal freeze-out of annihilations; they are due to a primordial asymmetry
between the number densities of particles and anti-particles. 
A scenario of this
type, ``Asymmetric Dark Matter'' (ADM), is also perfectly viable to explain the origin of the relic density of
dark matter particles.
%

One compelling idea is to have the dark matter asymmetry
arise due to a direct connection to the asymmetry in the baryons and 
leptons (see e.g. \cite{Kaplan:1991ah, farrar, ADM,Barr:1990ca,Thomas:1995ze,Kuzmin:1992up,Tytgat:2006wy,An:2009vq,Cai:2009ia,Kitano:2008tk,JMR}).
 In particular, one can imagine that the dark matter particle itself carries
baryon number and/or lepton number.  In that case the dark matter will share
in the asymmetry of regular matter in the early universe.  This general
scenario was proposed in \cite{ADM}, and it is the scenario that
we will be assuming throughout this paper.
One of its most
appealing qualities is that it leads generically to a prediction for the mass
of the dark matter particle;  since the asymmetry in the dark sector ends up
being roughly comparable to that in the baryons, the ratio of dark matter to
baryonic energy densities is given approximately by the ratio of the dark
matter mass $m_{\tmop{DM}}$ to the mass of the proton.  This leads to a
prediction for $m_{\tmop{DM}}$ of about $\CO(1-10)$ GeV depending on the
details of the model.  Here we are simply assuming that all interactions
which transfer the asymmetry between the dark and visible sectors have frozen
out of equilibrium by the time the temperature has dropped below
$m_{\tmop{DM}} .$\footnote{If the asymmetry
transfer freezes out below the DM mass, the densities of DM and
baryons are not related to each other in such a straightforward
way, and depend on the size of the DM annihilation cross-section, much
as in the standard thermal WIMP case. } 
It is an amusing coincidence that the prediction here for $m_{\tmop{DM}}$ is
fairly close to the weak scale, and thus could arise in models of electroweak
symmetry breaking, much as in the case of the standard weakly
interacting massive particle (WIMP).

The fundamental test of the WIMP hypothesis will eventually be an
issue of measuring the couplings of a given WIMP candidate, in order to check
whether the associated annihilation cross section would indeed have led to an
appropriate relic density.  Such a confirmation would yield convincing
evidence that the dark matter relic density did indeed have a thermal origin.
 Similarly, obtaining convincing evidence that the ADM mechanism is at play
may also be possible.  There would be two crucial ingredients which would be
important to confirm; first, one would like to check that the dark matter
particles do indeed carry a particular baryon number and/or lepton number, and
second, one would like to check that $m_{\tmop{DM}}$ is of an appropriate
size. 

A difficulty here is that, since we are requiring that the
operators responsible for carrying the baryon/lepton asymmetry between the
dark and visible sectors froze out at temperatures above $m_{\tmop{DM}}$,
these operators must be fairly suppressed.  On the other hand, it is
precisely these operators which would reveal the baryon/lepton number carrying
properties of the dark sector;  the requirement of early freeze-out implies
that signatures of these operators will in general be difficult to detect.

In this paper, we will discuss one reasonably generic signature
which the asymmetry transfering operators might have. 
If the dark matter carries baryon/lepton number, then decays or annhilations of this
particle into the standard model leptons and/or baryons 
may
 in general be made possible. 
Since electric charge is conserved and the DM particle
must be electrically neutral, the excess standard model baryon/lepton number
will necessarily show up either in the neutrino sector, or in equal
numbers of electrons and protons (or their anti-particles).  
Due to the requirements on
$m_{\tmop{DM}}$ in this scenario, neutrinos from DM decays 
would typically have energies
of order a few $\tmop{GeV}$, and could lead to a distinctive bump in cosmic
neutrino data in this energy range.  Moreover, due to the origin of the
signal, this bump would be associated with anti-neutrinos rather than
neutrinos (or vice versa), 
and this is a property which is potentially discernable at
the Super Kamiokande detector \cite{Ashie}, as well as MINOS \cite{MINOS} and
possible future detectors. 
 This feature would make an
observation of this type particularly compelling evidence in favor of the ADM
mechanism.

Unfortunately, in the case of operators leading to dark matter
annihilations, early freeze-out requirements constrain the event rates to be
considerably below the reach of both current and upcoming neutrino data. 
We relegate the details to appendix \ref{app:ann}.
As a result, we shall instead focus on operators leading to dark matter decays. 

As implied above, our most important limits will come from the Super Kamiokande
(Super-K) neutrino detector. The MINOS detector has 
about 20\% the fiducial volume of Super-K, though this is 
partly compensated for by its
ability to directly distinguish between neutrinos and anti-neutrions,
thereby cutting down on the atmospheric neutrino background in the ADM scenario. Although the IceCube \cite{ICECUBE} experiment now typically
sets more stringent limits on neutrino signals than Super-K, its low energy
threshold of $\sim 100 \tmop{GeV}$ is too high for our purposes.
We find that operators of dimension 6 leading to DM decays must be 
suppressed by at least $\sim 10^{12} - 10^{13}$ GeV,
due to existing neutrino constraints.  In many models of the early universe, 
the reheating temperature is required to be less than about the GUT scale,
and thus, upcoming improvements in the constraints will be probing into the remaining window where the operator responsible for decay is also capable of transfering the asymmetry between the SM and the DS.

The outline of this paper is as follows. In section 2, we review
the asymmetric dark matter (ADM) paradigm, and how one obtains 
quantitative predictions for the DM mass, focussing on a specific
choice for the operator responsible for DM decay to neutrinos. In
section 3, we discuss details of the production and observation
of neutrinos from DM decays, and the resulting constraints on 
the decay spectrum. We also discuss possibilities for distinguishing
neutrinos from anti-neutrinos at water-Cherenkov detectors and other
possible future detectors.  
 In section 4, we consider more general operators
which may lead to the decay of the DM particle. For each operator,
 our analysis gives a bound on the size of the scale suppressing
the interaction.
  In particular, we  consider
all possible interactions of dimension $\le 6$ coupling the DM sector
to a SM $(B-L)$-carrying gauge-singlet operator.   Finally, in section
5, we speculate on future directions and model-building issues.

%
%
%
%
%
%
%
%
%

\section{Asymmetric Dark Matter}
\label{sec:adm}

Asymmetric dark matter replaces the standard thermal history
of dark matter with one more closely analogous to that of baryons.
The relic density, rather than being fixed by the freeze-out of 
DM self-annihilation, is set by a small asymmetry between dark matter
particles and anti-particles, with all anti-particles eventually 
annihilating.  Various mechanisms for linking the dark matter
asymmetry to the baryon asymmetry of the universe have been
proposed in the literature \cite{Kaplan:1991ah,farrar,ADM,Barr:1990ca},
usually involving equilibrium processes that transfer any particle
asymmetry from the Standard Model (SM) to the dark sector (DS) and
vice versa.  We will focus on a specific class of models, where the
asymmetry is generated somehow (the details of which will not concern us
here) and transferred between the SM and DS by nonrenormalizable operators
of the form
\be
\Delta {\cal L} &=& \frac{{\cal O}_{\rm DS} {\cal O}_{\rm SM} } {\Lambda^{d-4}},
\label{eq:ops}
\ee
where ${\cal O}_{\rm SM},{\cal O}_{\rm DS}$ are gauge-invariant operators composed
solely of SM or DS fields respectively, and carrying equal and opposite 
nonzero baryon-minus-lepton ($B-L$) number.  

For the most part, we will leave model-building details aside and
work only with the operators (\ref{eq:ops}) below the scale at
which they are generated.  The spectrum and $(B-L)$-numbers of the
light DS states then determine which such operators can be generated.
We will focus on the lowest-dimension ($d\le 6$) operators of the form (\ref{eq:ops}).
It is of course not difficult for higher-dimension operators to be selected
by appropriate $(B-L)$ numbers, e.g. if $(B-L)$ of the DM particle
is $\ll 1$.
Also, note that we will consider some instances where some mechanism 
suppresses the asymmetry-transfering
operators beyond just powers of the scale where they
are generated, and dimensional analysis is somewhat incomplete in this case.
Still, classifying operators by dimension is useful anyway,
since it makes it easy to read off the lowest-dimension allowed operators
once the quantum numbers of the fields and any spurions are specified.  
In any case, the phenomenological signature of the decay is more sharply
peaked in energy when there are fewer decay products, making the signature
more distinctive; this in itself is
a reason to focus on operators of relatively low dimension.  

 Part of what makes
the mechanism we are considering interesting for a signature is that the 
gauge-invariant operator
$HL$ is renormalizable and appears in the majority of low-dimension
operators of the form (\ref{eq:ops}).  The operator  $HL$ has
been called the ``neutrino portal'' \cite{neutportal} 
since it can contract
with DS operators to allow dark matter decays to neutrinos. Neutrinos
from dark matter decays have the advantage over other decay products
that they are not electrically
charged, and thus their galactic propagation does not depend on 
difficult-to-determine astrophysics.  While neutrinos
are often produced in models as the result of dark matter decays, they
are rarely the dominant decay mode or the leading discovery channel.  
The only exception of which we are aware is the strongly coupled 
model in \cite{neutportal},
where the DS is a strongly coupled QCD analogue that can decay to
neutrinos and dark glueball states.  While this model is very
interesting, the ones that we are considering differ in several aspects.
First of all, there is the additional motivation for these models
based on the ADM mechanism.  Second, the DS in our case is perturbative
and perhaps simpler for determining the predictions of specific models.
Third, in the strongly coupled models, dark matter decays producing
charged baryons are significant enough to be a competitive
or leading signature compared to neutrinos in most of the parameter space
depending on one's assumptions about galactic propagation; essentially,
the neutrino constraints are stronger partly because they have the advantage
of not suffering as much from astrophysical uncertainties.  We will
see however that in a range of 
cases for ADM, charged lepton and baryon production 
is greatly suppressed.  This follows from the fact that ADM favors
the dark matter to be much lighter than in \cite{neutportal}.\footnote{
In fact, 
one could reasonably take the dark matter mass to be lighter in their
model as well in order to suppress the non-neutrino decays. The
reason \cite{neutportal} 
focused on 1 TeV$< m_{\rm DM} <$ 5 TeV was to get the correct
thermal relic density, but this could likely be modified with some 
additional model-building.} 
Finally, the assumption of ADM will
lead to some additional changes in the predictions of the model.
In particular,
ADM predicts that only anti-neutrinos and not neutrinos 
(or in some models, vice versa) ought to be
produced in dark matter decays.

We are interested in the $(B-L)$ ``portals'' of the SM, 
i.e. gauge-invariant operators that carry $(B-L)$ number. 
For this section and the next, we will focus on a simple illustrative example -
that of the operator $(LH)^2$, and an associated interaction
$\CO_{2\nu} \equiv \half \frac{X (LH)^2}{\Lambda^2}$, where $X$ is a complex
scalar DM particle. Note that for our purposes, the flavor indices in
this interaction are irrelevant, due to neutrino oscillations; all flavors
will be equally represented in the signal.
In later sections, we will consider more general operators for DM decay.

As usual, chemical equilibrium dictates
relations among the chemical potentials of each particle species
in the early universe
(see e.g. \cite{Yanagida,TurnerHarvey}), which we briefly review here.
The DM asymmetry then follows from the DM chemical potential, and determines
the DM particle mass.
In general, the number density asymmetry in a given species is $n - \bar{n}
  = 2 g \frac{T^2\mu}{6}$ for bosons and 
$g \frac{T^2\mu}{6}$ for fermions, where $g$ counts the number
of internal degrees of freedom (e.g., $g=1$ for a Weyl fermion).
We will calculate the constraints slightly differently from usual,
in a manner that emphasizes the dependence on the conserved charges
of the model.  This will allow us to discuss the implications
for the dark matter mass in terms of assumptions concerning the symmetries
of the dark sector couplings, rather than having to write down
specific dark sector interactions and analyze the resulting constraints.
This will be helpful later when we discuss more general models.

The point is that each interaction in chemical equilibrium imposes a linear
relation on the chemical potentials, of the form $\sum_i \mu_i=0$, where
the sum is over all the particles in the interaction. 
In the absence of
any conserved quantities, there are at least as many interaction terms
as there are fields with chemical potentials, 
and so the system is overdetermined and the chemical potentials vanish.
 However, for each $U(1)$ conserved charge,
 it will also be true that $\sum_i q_i=0$, where
$q_i$ is the charge of the i-th particle in the interaction.
  Therefore, the choice of
$\mu_i = c q_i$ leads to a non-vanishing solution for the chemical
potentials, where $c$ is arbitrary.  Since the constraint
equations are linear, the general solution will be a sum over
the solutions for each abelian symmetry.


In the Standard Model, the only two flavor-universal,\footnote{For 
simplicity, we assume that the asymmetry is generated
to be flavor-universal. Alternatively,
if the DM couples to all lepton flavors, then these interactions
will enforce flavor-universal asymmetries in the SM regardless
of the flavor-independence of the original asymmetry. 
At any rate, this does not qualitatively affect our results.}
linearly independent,
anomaly-free
$U(1)$s are hypercharge and $(B-L)$. Labeling the two corresponding
coefficients $c_Y$ and $c_{B-L}$, we have
\be
&&\mu_H = \half c_Y, \ \ \ \ \ \ \ \ \ \ \ \ \ \ \ \mu_l = -\half c_Y - c_{B-L}, \ \ \ \
  \mu_e = -c_Y - c_{B-L} ,\nn \\
&& \mu_q = \frac{1}{6} c_Y + \frac{1}{3} c_{B-L}, \ \ \
   \mu_u = \frac{2}{3} c_Y + \frac{1}{3} c_{B-L}, \  \ \ 
   \mu_d = -\frac{1}{3} c_Y + \frac{1}{3} c_{B-L}.
\label{eq:chempots}
\ee
One may easily check that $c_Y$ and $c_{B-L}$ cancel out of
the constraint imposed by any Standard Model interaction; for
instance, $Q U^c H$ imposes $\mu_q - \mu_u + \mu_H=0$, which is
automatically satisfied by (\ref{eq:chempots}).  In general,
there may be accidental symmetries that can introduce additional
parameters. However, the accidental symmetries $B$ and $L$
separately are efficiently violated above the electroweak scale
by sphaleron processes \cite{KRS}.   

Additionally, the total hypercharge must vanish, 
\be
&&N_f\left(\mu_q + 2\mu_u - \mu_d - \mu_l -\mu_e\right) +2 \mu_H 
 = 11c_Y + 8c_{B-L} = 0,
\ee
leading to a single remaining free parameter required to fix all of the
chemical potentials, which must be set
by initial conditions. It is convenient
to parameterize this in terms of the total baryon asymmetry $B = N_f (2\mu_q
+\mu_u + \mu_d)$, where
$n_B - \bar{n}_B \equiv \frac{T^2 B}{6}$, leading to
\be
 c_{B-L}  = \frac{11}{28} B, &&  c_Y = -\frac{2}{7} B \ .
\ee
For ${\cal O}_{2\nu}$, the conserved quantum numbers of
the dark matter particle are completely determined by the interaction, and 
one obtains a definite
prediction for the dark matter particle chemical potential and 
mass:  $\mu_X = 2c_{B-L} = \frac{11}{14}B$ and 
$m_X = \frac{\Omega_{\rm DM} }{\Omega_b} \frac{ B }{2 \mu_X} m_p 
\approx 3$ GeV.\footnote{
This prediction for the mass is modified by $\CO(20\%)$ if the 
asymmetry-transferring
processes freeze out below the electroweak scale \cite{ADM}. }
This is representative of the typical size of mass favored by ADM.  The 
decay spectrum into a particular neutrino flavor is in this case effectively 
$\frac{dN}{dE} = \frac{2}{3} \delta(E_n-
m_{\rm DM}/2)$. 
This is independent of
the relative branching ratios of decays to different flavors, due to
neutrino oscillations.

\section{Neutrino Flux}

\subsection{Halo Flux}

The primary source of dark matter decays is from the galactic halo.
The flux of neutrinos of a given flavor detected at angle $\psi$ is
\be
\frac{d \Phi}{d E} &=& \frac{\Gamma_{DM}}{4\pi m_{\rm DM}} \frac{dN}{dE} 
R_{sc} \rho_{sc} \Delta \Omega {\cal J}_{\Delta \Omega}(\cos \psi),
\ee
where $dN/dE$ is the spectrum of decays into a specific flavor, 
$\Gamma_{DM}$ is the
decay width to all neutrino flavors, and $m_{\rm DM}$ is the dark matter mass. 
Here we have defined the average flux integral ${\cal J}_{\Delta \Omega}$
as follows. First, we take the conventional definition for
the line-of-sight (LOS) flux integral ${\cal J}$:   
if the direction the telescope is pointed in
makes an angle $\psi$ with the direction toward the galactic center (GC),
then 
\be
{\cal J}(\cos\psi) &=& \frac{1}{R_{sc} \rho_{sc}} \int_0^{l_{\rm max} }
  \rho(\sqrt{R_{sc}^2 - 2x R_{sc} \cos \psi + x^2}) dx,
\ee
where the upper limit $l_{\rm max} = \sqrt{R_{mw}^2 - \sin^2 \psi R_{sc}^2}
 + R_{sc} \cos \psi$ depends on the size of the Milky Way halo, which we
take to be $R_{mw} = 34$kpc,
and $R_{sc}=8.5$kpc, $\rho_{sc} = 0.3$GeV$/$cm$^3$ are the galactic 
radius and density at the solar circle in the NFW halo profile, respectively.  
For an effective detector resolution
of $\theta$, we then take the average of 
${\cal J}(\cos\psi)$ over a cone of half-angle $\theta$ (i.e. solid angle
$\Delta \Omega = 2\pi (1-\cos \theta)$):
\be
{\cal J}_{\Delta \Omega}(\cos\psi) &=&
   \frac{1}{ \Delta \Omega} \int_0^\theta \sin\theta' d\theta' 
  \int_0^{2\pi} d \phi
{\cal J}(\hat{r} \cdot \hat{r}_{\rm GC} ),
\ee
where $\hat{r} \cdot \hat{r}_{\rm GC} = \sin \psi \sin \theta' \cos \phi
+ \cos \psi \cos \theta'$.
${\cal J}_{\Delta \Omega}$
  is at a maximum in the direction of the GC, at $\psi=0$.

The size of the signal from dark matter decays to neutrinos depends
somewhat on the galaxy halo profile.  To give a sense of this
dependence, we will compare the size of the signal for four
different halos: the Einasto profile
\be
&&\rho_{\rm Ein} = \rho_0
 \exp\left[ -2\left( \left( r/r_s \right)^{\alpha} -1\right)/\alpha 
\right]
\ee
 and
the NFW, Moore, and Kravtsov profiles, which have the parameterization
\be
&& \rho = \rho_0 \left( r/r_s \right)^{-\gamma}
   \left( 1 + (r/r_s)^\alpha \right)^{(\gamma-\beta)/\alpha}.
\ee
The values of the parameters for these models are given in Table
1. 
The largest variation in the halo profiles is nearest the galactic center.

For dark matter decay,  unlike the case
for dark matter annihilation, the size of the signal near the galactic center
is not strongly dependent on the halo profile.  
In Table 1, we show ${\cal J}_{\Delta \Omega}(1)$ (i.e.,
pointed directed at the galactic center) for cones of half-angle
$10^\circ, 30^\circ$, and $180^\circ$.   As one can see, 
the variation is fairly modest
and less than a factor of 2 in even the most extreme example.

\begin{table}
\begin{center}
\begin{tabular}{ccccccccc}
\hline
 & ${\cal J}_{\theta=10^\circ}(1)$ & ${\cal J}_{\theta=30^\circ}(1)$ & 
${\cal J}_{\theta=180^\circ}(1)$ 
& $r_s($kpc$)$ & $\rho_0$(GeV$/$cm$^3$) & $\alpha$ & $\beta$ & $\gamma$\\
\hline
Einasto & 12.6 & 6.7 & 1.8 & 20 & 0.06 & 0.17 & n/a & n/a\\
Moore & 13.9 & 6.5 & 1.8 & 28 & 0.053 & 1.5 & 3 & 1.5\\
NFW & 10.2 & 6.0 & 1.9 & 20 & 0.26 & 1 & 3 & 1\\
Kravstov & 6.8 & 5.5 & 2.1 & 10 & 0.703 & 2 & 3 & 0.4 \\
\hline
\end{tabular}
\caption{${\cal J}_{\Delta \Omega}(1)$ for various halo models and angular
averages ($\Delta \Omega = 2 \pi (1- \cos \theta))$.  }
\end{center}
\label{tab:halos}
\end{table}

\subsection{Cosmic Flux}

The contribution to the neutrino signal from cosmic dark matter
decays is small compared to the halo contribution, but for
completeness, we will review this here (see e.g. \cite{Beacom} for
more details).  The cosmic flux signal is given by
\be
\frac{d \Phi}{dE} &=& \frac{\Gamma_{DM}}{4 \pi m_{\rm DM}} 
\frac{ \Omega_{\rm DM} \rho_c}{H_0} 
\int_0^\infty \frac{dN(E')}{dE'}\frac{dz}{h(z)}, 
\ee
where $\rho_c$ is the universe critical density, $H_0$ is the
Hubble constant today, $\Omega_{\rm DM} = 0.22$, $E'\equiv (1+z)E$ is
the redshifted energy, and
$h(z) = [(1+z)^3 \Omega_m + \Omega_\Lambda]^{1/2}$. For comparison
with the halo flux, we may define an effective ${\cal J}_{4\pi}$ for
the cosmic flux:
\be
{\cal J}_{4\pi, {\rm Eff}} &\equiv & \frac{\frac{\Omega_{\rm DM} \rho_c}{H_0} 
\int_0^\infty {dz \over (1+z) h(z)} }{4 \pi R_{sc} \rho_{sc}} = 0.16 \ .
\ee
(The additional $(1+z)$ in the denominator is from the integration over
energy.)  This is negligible even compared to the full $4\pi$ average
${\cal J}$ in any of the halo models, and we will ignore it from now on.

\subsection{Neutrino Constraint}

\begin{figure}[t!]
\begin{center}
\includegraphics[width=0.7\textwidth]{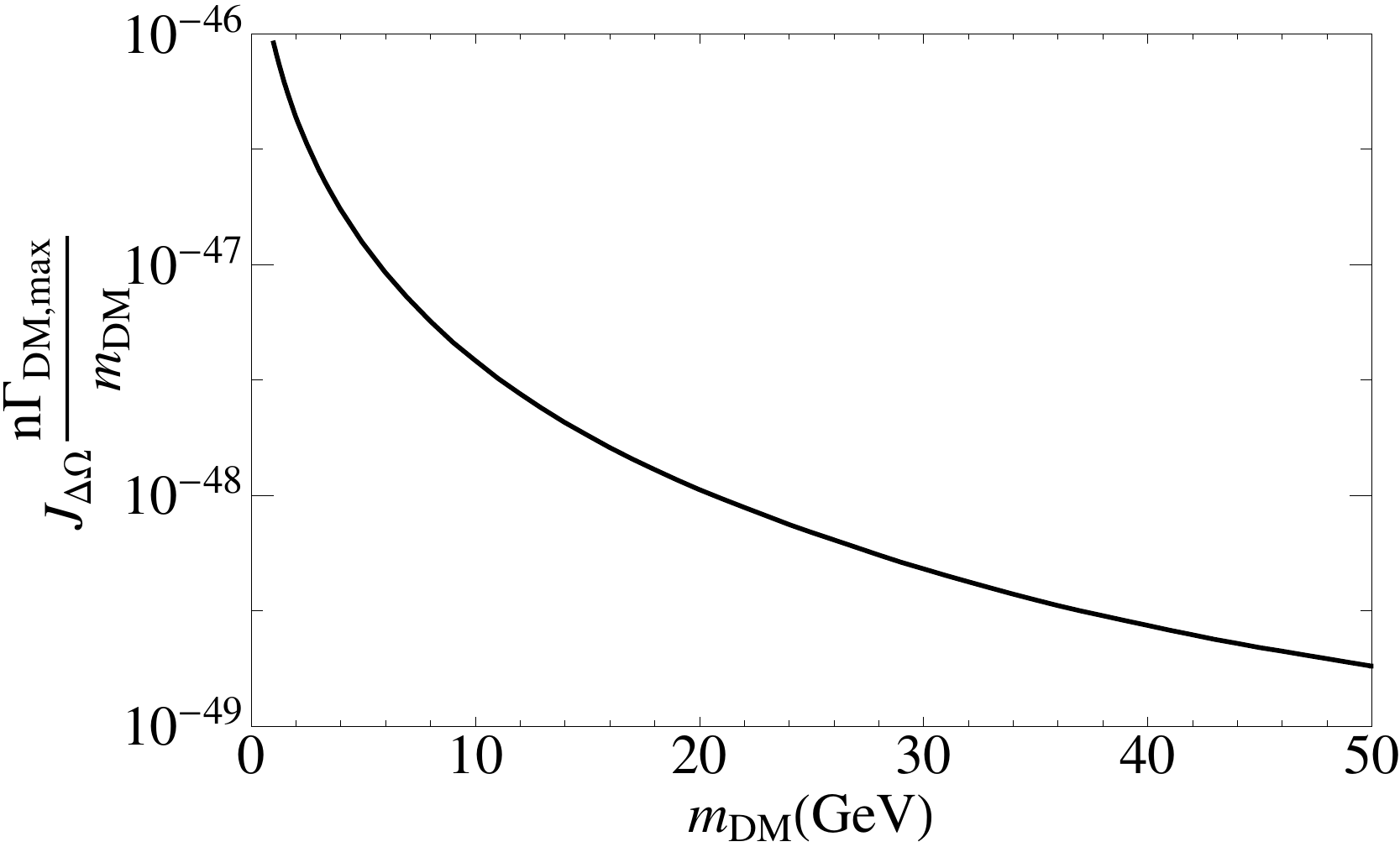}
\caption{ Bound on total decay width to all flavors of neutrinos for
a monochromatic spectrum, from demanding that
the signal that would be observed at Super-K not be greater than
the atmospheric muon-neutrino $(\nu_\mu + \bar{\nu}_\mu)$ background. 
The bound depends on the solid angle $\Delta \Omega$ of observed sky
around the galactic center, and on
the number $n$ of neutrinos produced in each decay. ADM typically
favors masses in the lower range of this plot. }
\label{fig:decaybound}
\end{center}
\end{figure}

To derive our constraints, we will follow the analysis of
\cite{Beacom}. To be conservative, we demand that the predicted muon-neutrino 
flux
at Super Kamiokande from dark matter decays be less than the 
background from atmospheric $(\nu_\mu + \bar{\nu}_\mu )$ neutrinos.  
For two-body decays, neutrinos
from the halo decays will be nearly monochromatic.
  Thus, following
\cite{Beacom}, we choose
an energy bin of width $\Delta \log_{10} E = 0.3$, which is
about the width of the bins Super-K uses in their own analyses \cite{Ashie}, 
centered on $E= m_{\rm DM}/2$,
and we demand that the total predicted signal in this bin be less than
the flux of atmospheric muon-neutrinos.
Note that for multi-body decays, this would contain only some fraction of the
total decay spectrum, making the bound on the decay rate
correspondingly weaker.  Since the uncertainty
in the background is only $\sim 10$ percent, one expects a dedicated analysis of
Super-K data to improve the $2\sigma$ 
bound on the decay width by roughly a factor
of five; such an analysis is currently underway \cite{Ed}.  
For the background rate, we use the flux calculated
by Monte Carlo in \cite{Honda:2006qj}, and take the average over the full sky.
The resulting constraint is shown in Fig. \ref{fig:decaybound}.
We have explicitly factored out the model-dependent number $n$ of
neutrinos produced in the decay, and the angular integral factor
${\cal J}_{\Delta \Omega}$. 
The above bound translates into a limit on the scale $\Lambda$ appearing
in $\CO_{2\nu} = 
\half \frac{X (HL)^2}{\Lambda^2}$ of $\Lambda > 6 \times 10^{13}$ GeV,
taking for example ${\cal J}_{\theta=30^\circ}$.
This operator freezes out at temperatures of order $T_{\rm fr} \sim 
(\Lambda^4/M_{\rm pl})^{1/3}
\gtrsim 10^{12}$ GeV.  Assuming that this operator is responsible for the
asymmetry transfer and is still present in the 
effective theory at high temperatures $T_{\rm rh}$ where reheating occurs, 
$\Lambda$
must be less than $(T_{\rm rh}^3 M_{\rm pl})^{1/4}$ in order for the
asymmetry transfer to occur at all.
If $T_{\rm rh}$ is $M_{\rm GUT}$ or less, which is true of most models, then the
remaining window for $\Lambda$ is $6 \times 10^{13} \textrm{GeV} \lesssim
\Lambda \lesssim 5\times  10^{16} \textrm{GeV}$.  
Thus, future improvement in the
neutrino constraint will probe interesting regions of the parameter
space, based not only on the closeness of $\Lambda$ to the GUT scale but
also due to expected limitations on the freeze-out temperature. 

\subsection{Distinguishing $\nu$s from $\bar{\nu}$s}

While seeing a feature in the neutrino spectrum at $O(1-10$ GeV$)$ might
give suggestive evidence in favor of the ADM picture, a verification of the
lepton number violating origin of the signal would be crucial to making this
evidence truly compelling.  
As
we will discuss further in section \ref{sec:othermodels}, 
the value of the lepton number carried by the dark
matter particle is model dependant, and may be either positive or negative.
With the relatively low dimension operators we consider, this implies that
the neutrino signal will be composed entirely of either neutrinos or
anti-neutrinos.  An important question is then whether or not this feature is
distinguishable in present detectors, or at future neutrino
experiments.

If a signal were to be seen at MINOS despite its smaller fiducial
volume than Super-K (4 kton vs. 20 kton), it would trivially
be able to test the neutrino vs. anti-neutrino composition of the 
excess due its magnetic field (through the charge of the same-family lepton
produced in interactions).  In fact, since the detected background
of atmospheric neutrinos 
is about 2 times larger than for anti-neutrinos, MINOS would have
the advantage that it could cut down on its background relative to 
an ADM anti-neutrino signal.  
At water Cherenkov detectors such as Super-K, there are several possible
avenues for separating neutrinos from anti-neutrinos on a statistical basis,
even though there is no direct way of measuring the signs of particle
charges.  For dark matter masses below about $5$ GeV, the most promising
one involves looking for neutrino interactions in the detector whose final
states contain one extra muon, 
beyond the number expected from the flavor of
the original neutrino.  Such muons 
are produced as decay products
of charged pions, and, though they are typically produced with little kinetic
energy, they are easily identified by the appearance of a decay electron one
muon lifetime after the initial event, i.e., a ``muon-decay electron''.   
The dominant neutrino interactions
(for, e.g., electron neutrinos)
leading to charged pion production are

\begin{eqnarray*}
  \nu_e + P \rightarrow P + e^- + \pi^+,  &  & \\
  \nu_e + N \rightarrow N + e^- + \pi^+ , &  & \\
  \bar{\nu}_e + P \rightarrow P + e^+ + \pi^- , &  & \\
  \bar{\nu}_e + N \rightarrow N + e^+ + \pi^-.  &  & 
\end{eqnarray*}
We thus see that neutrinos tend to lead to the production of positively charged
$\pi$'s, while anti-neutrinos tend to lead to the production of negatively charged
$\pi$'s.  The key point now is that, before having a chance to decay, a $\pi^-$
traveling through the detector with around a  GeV of energy tends to become
absorbed by a proton.  Coulomb repulsion of
$\pi^+$'s, on the other hand, suppresses inelastic scattering of these
particles, so that,
after stopping in the detector, they typically are successfully able to decay.  For this reason,
looking for events with an extra muon-decay electron 
in the final state at energies around a
GeV tends to efficiently select about 10 times as many  neutrino as
anti-neutrino events \cite{Ed}.  In general, since the total cross
section for detection of neutrinos is about a factor 2 larger than for anti-neutrinos
to begin with,
the result is a net  increased sensitivity to anti-neutrinos by about a factor of 5.  
Unfortunately, as the pion energy becomes much higher than a GeV 
(corresponding to DM masses much higher than roughly 5-10 GeV), both
$\pi^+$'s and $\pi^-$'s tend to scatter inelastically before decaying in the detector, and
the method is no longer
successful.

A few other methods may be used to distinguish neutrinos from anti-neutrinos
at water Cherenkov detectors, but it is unclear if Super Kamiokande would have
sufficient statistics to make use of them if they were to find a bump in the
neutrino spectrum at the energies suggested by ADM.  These might thus only
be useful at future detectors with larger volumes, such as Hyper Kamiokande,
or one at DUSEL at the Homestake mine \cite{HyperK, DUSEL}.  In \cite{SolarAnti},
differences in angular distributions for final state leptons in neutrino
vs anti-neutrino interactions were used to look for anti-neutrinos coming
from the sun.  In the ADM case, one could similarly make use of the increased
flux that a signal would lead to coming from the galactic center.  Finally,
inelastic scattering of neutrinos via the interactions

\begin{eqnarray*}
  {\nu}_e + N \rightarrow P + e^-,  &  & \\
  {\nu}_{\mu} + N \rightarrow P + {\mu}^-,  &  &
\end{eqnarray*}
allows for very efficient selection of ${\nu}$ over $\bar{{\nu}}$ events, but
only if one is able to identify the proton in the final state.  This has been
shown to be possible for protons with around a GeV of energy by making use
of the short length of their Cherenkov tracks,  but the required cuts are
currently rather inefficient \cite{ProtonID}.

Finally, we note that, if INO (India Neutrino Observatory) were to be built,
its magnetic field would allow for trivial separation of ${\nu}$ from
$\bar{{\nu}}$ events \cite{INO}.  Similarly a possible future liquid argon detector such as
GLACIER \cite{GLACIER} or one at DUSEL \cite{LANNDD} would be able to efficiently identify protons
and would also lead to efficient ${\nu}$/$\bar{{\nu}}$ tagging.

\section{Other Models}
\label{sec:othermodels}

The low dimension ($\le 5$) $(B-L)$ portals in the Standard Model 
are $ LH, (LH)^2, LLE^c, U^c U^c D^c$, and $LQD^c$.
The portal is opened, so to speak, by operators such as (\ref{eq:ops}).
Denoting bosonic dark sector states by $X$s and fermionic dark sector
 states by $\psi$s, we may list all such operators from lowest dimension
to highest dimension up to dimension $d=6$ (suppressing flavor indices)
\footnote{Our convention for $H$ is that $v=$175 GeV.}:
\be
d=4: \ \ {\cal O}_1 &=& \psi HL \nn\\
d=5: \ \ {\cal O}_2 &=& X \psi LH\nn\\
d=6: \ \ {\cal O}_3 &=& \psi L L E^c \nn\\
{\cal O}_4 &=& \psi L Q D^c \nn\\
{\cal O}_5 &=& \psi U^c D^c D^c \nn\\
{\cal O}_6 &=& X_1 X_2 \psi L H \nn\\
{\cal O}_{2\nu} &=& \half X (LH)^2 
\label{eq:oplist}
\ee

In order for ${\cal O}_1$ to lead to a decay of $\psi$, an $H$ must be
produced in the interaction. 
Since the dark matter in the cases we consider will always be lighter
than the Higgs, we may as well just integrate it out. Thus, for the sake
of considering signatures, ${\cal O}_1$ is just a special case of
${\cal O}_3, {\cal O}_4$.  At any rate, 
${\cal O}_1,{\cal O}_3,{\cal O}_4$,
and ${\cal O}_5$ are all models
with $(B-L)[\psi]=1$, which we will not consider further since neutrinos
are not the dominant decay signature.  However, there is nothing in principle
wrong with $(B-L)[\psi]=1$, provided the marginal operator ${\cal O}_1$
is sufficiently suppressed to evade particle physics constraints.\footnote{
Obviously, if the SM/DS interactions violate baryon number {\em and} lepton 
number, one must take care that any operators leading to proton decay
are sufficiently suppressed.  None of the operators we will be interested
in carry baryon number, so this is a question for the UV completion, which
is easily addressed.  For instance, a sufficient but not necessary condition
is that baryon and lepton number are accidental symmetries in the DS and
SM below the GUT scale. }
${\cal O}_5$ has been studied in the context of baryon number violation
in SUSY theories (see \cite{Sjostrand:2002ip} and references therein)
and is predicted to result in production
of a slow-moving (momentum $ \sim $ GeV) baryon from DM decays.

It might seem based on the ratio of $\Omega_b/\Omega_{\rm DM}$ 
that the mass is restricted to $\sim 5$ GeV. However, 
a simple example illustrates that $m_{\rm DM} = \CO(1-50)$ GeV
may also be obtained fairly easily.
The reason is that for many of the operators in (\ref{eq:oplist}),
the prediction for the mass depends on the model-dependent
quantum numbers and spectrum of the dark sector particles. 
Our example is ${\cal O}_2$ with $X$ being the DM particle,
and supposing that  there are no other
conserved abelian symmetries under which $X$ and $\psi$ are charged.
From our analysis in section \ref{sec:adm} of chemical equilibrium,
this leads to a prediction for the dark matter mass of
$m_X \approx 6.4 (-l_X)^{-1}$ GeV, where
$L(X)=l_X, L(\psi)=-1-l_X$.\footnote{If $l_X$ is positive
instead of negative, then there are more $X^*$s than $X$s at late
times, and $X$ particles are anti-dark matter.}  
Note that this argument applies generally to bosonic DM
with lepton number $l_X$; for fermion DM,
the mass prediction is $m_X \approx 13 (-l_X)^{-1}$ GeV.  
Such DM masses can quite reasonably be between say $\sim$ 1 and 50 GeV.

The spectrum of observed neutrinos will be somewhat model-dependent.
For illustration, we may continue to consider ${\cal O}_2$ and note two 
different kinematic
possibilities. 
First,  if $\psi$ is lighter than $X$, then $X$ can decay to $\psi\nu$.\footnote{In this
case $\psi$ must have an appropriate mass so that it
does not contribute significantly to the late-time density of the universe.
Alternatively, $\psi$ may contribute to the dark matter density today, but then
the prediction for the mass of $X$ must be correspondingly diminished.}
Alternatively, if $\psi$ is heavier, $X$ may still be able to decay
through a virtual $\psi$.  In fact, if $\psi$
decays to additional light DS states are allowed, then we can just integrate out $\psi$ and
describe the decay by an operator of the form of ${\cal O}_6$. The former possibility is a two-body
decay and thus leads to a monochromatic spectrum, whereas the second
is multi-body and the spectrum is somewhat smeared out in energy, and the bound on $\Gamma_{\rm DM}$ becomes slightly weaker.\footnote{ In
general, we ought to
have a reason why the $X$ decay to light DS states is sufficiently
slow.  A simple reason is to use $(B-L)$ conservation and have $|l_X|$
smaller than the lepton number of any lighter DS states.
For example, if $l_X=-1/3$
and $\psi$ is the only leptonic dark sector state lighter than $X$, then
$X$ can decay only through suppressed higher dimensional operators 
like ${\cal O}_2$ that involve the Standard Model. This is essentially
a dark analogue of invoking baryon triality to forbid proton decay. 
}


The decay spectra for several operators are listed in Table 2.\footnote{We take $\psi$ to be a Dirac fermion when calculating 
decay rates. If $\psi$ is the DM, then this is necessary
anyway in order to give $\psi$ a DM-number conserving mass term.}  Using them,
the bound on $\Gamma_{DM}/m_{\rm DM}$ in Fig. \ref{fig:decaybound} 
is then easily translated into
a bound on the coefficients of the ADM operators, once the DM mass is
specified.
\begin{table}
\begin{center}
\begin{tabular}{c|c|l|l}
\hline
$\Delta {\cal L}$ & Decay & $\Gamma_{DM} \frac{dN}{dE_\nu}$ & $\Lambda_{\rm min}$ (GeV) \\
\hline
${\cal O}_2/\Lambda$ & $X\rightarrow \psi \nu$ & $\frac{m_{\rm DM}}{(3)(32\pi)}
\frac{v^2}{\Lambda^2} \delta(E_\nu-m_{\rm DM}/2)$ & $8\times 10^{24}$ \\
${\cal O}_{2\nu}/\Lambda^2$ & $X\rightarrow \nu\nu$ &
$\frac{m_{\rm DM}}{(3)(8\pi)} \frac{v^4}{\Lambda^4} 
\delta(E_\nu-m_{\rm DM}/2)$  &  $6 \times 10^{13}$ \\
${\cal O}_6/\Lambda^2$ & $X_1 \rightarrow X_2\psi\nu$ & 
$\frac{v^2}{(3)(4\pi)^3 \Lambda^4} 
E_\nu^2 \Theta\left( \frac{m_{\rm DM}}{2} - E_\nu \right)$ & $10^{12}$  \\
${\cal O}_6/\Lambda^2$ & $\psi \rightarrow X_1 X_2\nu$ & 
$\frac{v^2}{(3)(4\pi)^3 \Lambda^4} 
E_\nu^2 \Theta\left( \frac{m_{\rm DM}}{2} - E_\nu \right)$  &  $10^{12}$ \\
$ \frac{g (\psi H L)(\chi \chi^c)}{\Lambda m_X^2} $ & $\psi \rightarrow
\chi \chi^c \nu$ & 
$ \frac{2 g^2 v^2m_{\rm DM}}{(3)(4\pi)^3 \Lambda^2 m_X^4} 
E_\nu^2 (\frac{m_{\rm DM}}{2} - E_\nu) \Theta\left( \frac{m_{\rm DM}}{2} - E_\nu \right)$ &
$1.4 g \left(\frac{10 \textrm{GeV}}{m_X}\right)^2 10^{22}$ \\
\hline
\end{tabular}
\label{tab:spectra}
\caption{Decay rates and spectra for a range of effective operators
responsible for the decay of a dark matter particle with mass $m_{\rm DM}$. 
 The specific decay process of the dark
matter particle is shown. $X$s are used for bosonic DM particles
and $\psi,\chi$s are used for fermionic ones. $E_\nu$ is the energy
of the neutrino, and $v\equiv$ 175 GeV. In all cases, we have
taken $m_{\rm DM}= 3$ GeV and ${\cal J}_{\Delta \Omega} = {\cal J}_{\theta=30^\circ }$
for calculating $\Lambda_{\rm min}$.}
\end{center}
\end{table}
%
%
In the table is indicated the operator that gives rise to the decay,
as well as the specific decay process that is assumed.  
For instance, the third and fourth rows give the decay spectra for
the operator ${\cal O}_6$, in the case of  $X_1$ (i.e. bosonic) dark matter
 and in the case of $\psi$ (i.e. fermionic) dark matter, respectively. 
The final row presents the spectrum in the scenario where 
the asymmetry is transferred by the
operator ${\cal O}_2$ but the dark matter particle $\psi$ is
lighter than $X$,
which is produced off-shell and decays to two Weyl fermions
$\chi, \chi^c$ through the interaction $g X^* \chi \chi^c$, 
generating ${\cal L} \supset \frac{g}{\Lambda m_X^2}(\psi HL)(\chi \chi^c)$.
In all but the last example above, 
the spectrum still has a fairly sharp feature at $E_\nu = m_{\rm DM}/2$, and so
is nearly as easily detected as a two-body decay.  In fact,
the fraction of events that fall in a bin of size $\Delta \log_{10} E=
0.3$ is 87\% for the $\CO_6$ spectrum and 71\% for the last example.
We have included this in the calculation of $\Lambda_{\rm min}$ in 
Table 2, and in all cases we have taken $m_{\rm DM}=$ 3 GeV for
the sake of concreteness. 

While Table 2 shows that there is clearly a lot of model-dependence in
the bound on the UV scale $\Lambda$, we will note two points.  
The first is that for the dimension 5 operator
$\CO_2$, the scale is forced to be much larger than the Planck scale, and
so is ruled out in the absence of additional structure in the DS.
We will discuss this further in the next section.  The second
point is that for dimension 6 operators, the bound on $\Lambda_{\rm min}$
is about $10^{12} - 10^{13}$ GeV, and thus the remaining window of
open parameter space under the GUT scale is only about 3-4 
orders of magnitude in $\Lambda$.


There will also be some small branching ratio for dark matter to decay 
through an off-shell Higgs to charged Standard Model particles; however,
this will be very suppressed since the dark matter is quite light. 
Details of the comparison between a signal from neutrinos vs. from charged 
particles are given in appendix \ref{app:dom}, with
the conclusion that decays to charged particles are negligible for the
operators we are considering.

\section{Discussion}


Asymmetric dark matter is a compelling and simple framework, alternative
to the standard thermal WIMP paradigm.  In particular, ADM favors 
lighter DM masses around $\CO(1-10)$ GeV, which have received a 
recent boost in interest due to hints of a signal coming
from the CoGeNT direct detection experiment \cite{COGENT}, as well
as possible explanations for the DAMA anomaly \cite{DAMA,DAMAChannel,
 Feldstein, Fitzpatrick, Hooper, Chang, Petriello, Gondolo, WIMPless}. 
The phenomenological implications
can differ in significant ways from those of a standard thermal WIMP,
and we have noted one such possibility here, in which the DM particle carries
lepton number and decays dominantly to anti-neutrinos.  Such a scenario
is motivated as a reasonably generic consequence of a mechanism for 
transferring the SM baryon asymmetry
to the DS by virtue of the lepton number of the DM particle \cite{ADM}.

There are various model-building issues worth exploring in ADM.
One would like to specify the mechanism that generates the asymmetry
in the first place, and how the symmetric component is to be removed.
Furthermore, one needs to generate a mass for the dark matter particle
with the appropriate size to give the correct relic density.  The fact
that the typical masses required are ${\cal O}(1-10$ GeV$)$ suggests
a common origin with the electroweak scale, suppressed by additional
loop factors or smaller couplings. If the dark matter is a scalar,
the mass also needs to be protected from radiative corrections.
This is clearly related to the issue of forbidding or suppressing the marginal 
coupling $|X|^2 |H|^2$ to the Higgs.  Also, since conservation of 
lepton number is protecting the lifetime of the dark matter particle, it may
not always be straightforward to take advantage of the see-saw mechanism for
Standard Model neutrino masses.  

We will comment briefly on how these issues may be addressed in one
example using mechanisms already suggested in the literature. 
As noted in \cite{ADM}, the operator ${\cal O}_2$ is
attractive since it may easily 
be UV-completed to a model with Standard Model Majorana
neutrino masses arising from the see-saw mechanism.  
For example, we may add two $SU(2)$ doublet fermions $d,d^c$ with 
$L=1+l_X, -1-l_X$ and hypercharge $Y=\half, -\half$ respectively, and take 
\be
\Delta {\cal L} &=& \lambda \psi d H + \lambda' L d^c X + m_d d d^c.
\label{eq:simpmodel}
\ee
The right-handed neutrino masses arise from a scalar vev $\phi$ with
$L=+2$.  Thus no renormalizable couplings of $\phi$ to the
fields $X,\psi,d,d^c$ are allowed as long as $ -\frac{2}{3} < l_X < 0 $,
 and lepton number can then easily be an accidental symmetry in the dark sector.  
Note, however, that in this particular 
example, the operator ${\cal O}_2$ is parametrically suppressed
only by the scale $m_d$, which would be constrained by observations 
to be above the
Planck scale according to Table 2 if the decay of the dark matter particle $X \rightarrow \psi \nu$
were kinematically accessible.  On the other hand, simple extensions may suppress
${\cal O}_2$ without requiring such high scales.  For example, 
consider a model 
with an extra gauged $U(1)$ broken by a scalar vev $\langle \Phi
\rangle$, and two copies $d_i, d_i^c$ of the doublet fields, 
where the non-zero $U(1)$ charges of the fields are
as follows:

\begin{center}
\begin{tabular}{c|c}
 & U(1) \\
\hline
$X$, $d_2$ & 1 \\
$\Phi$, $d_2^c$ & -1
\end{tabular}
\end{center}

%
%
\noindent
and 0 for the remaining fields. 
Then symmetry forces the lagrangian to be of the form
\be
\Delta {\cal L} &=& \lambda \psi d_1 H + \lambda' L d^c_2 X + 
 m_{d,i} d_i d_i^c + c \Phi d_2 d_1^c,
\ee
which upon integrating out the $d_i$ fields leads to ${\cal O}_2$ having a coefficient whose parametric suppresion is
$\frac{c \lambda \lambda' \langle \Phi \rangle }{m_d^2}$.

So far, in our discussions, the DM mass has been put in by hand.  However, if the baryon-to-dark
matter density ratio is truly to be explained, then $m_{\rm DM}$
must be dynamically generated at the correct scale.
As noted in \cite{Morrissey:2009ur}, 
a mass of ${\cal O}(1-10$ GeV$)$ arises quite 
generically in models of spontaneously
broken supersymmetry from gravity-mediated effects, provided that the MSSM
masses are generated by gauge mediation with a messenger scale at
$M \sim 10^{13}$ GeV. It would be more satisfying, though, to have a 
model where the mechanism that generates the DM mass is more predictive, and 
does not depend on a mass parameter that is free to vary over several orders
of magnitude.   Supersymmetry has the advantage that 
in theories with large
$\tan \beta\sim 20$, allowing a superpotential 
term of the form $W \supset X H_d S$,
where $S$ is a doublet,
generates a mass for $X$ at the appropriate size.
Furthermore, additional contributions to the dangerous 
marginal term $|X|^2 |H|^2$ can come only from superpotential terms,
which may be easily controlled.
    
There are many ways the asymmetry can be generated;
for an incomplete list see \cite{KolbTurner}.  It is perhaps
worth noting that the asymmetry does not have to originate in the SM and get
transferred to the DS but could instead 
originate due to new sources of CP violation in the DS and then be
transferred in the other direction. 

Finally, there are other possible ways that neutrino signatures of ADM 
could appear.  One potentially interesting direction 
currently under investigation 
involves neutrinos coming from DM annihilation in the sun.  ADM is
very interesting in that its annihilation cross-section
is not a priori related to the cross-section for capture in
the sun.  Indeed, because of conservation of lepton number,
the scattering process will never contribute radiatively to
the annihilation process, and thus
there is no theoretical barrier to taking the two cross-sections
to be quite different.\footnote{  It is conceivable that an 
additional potentially significant
boost in the solar signal could come from exponential growth
of the DM occupation in the sun due to WIMP-WIMP scattering
\cite{zentner}.}

\section*{Acknowledgments}
We would especially like to thank Ed Kearns and Jennifer Raaf for many useful
discussions concerning the Super Kamiokande experiment and neutrino
detection. We would like to thank E. Katz for useful discussions, and 
M. Schmaltz for helpful discussions and comments on the manuscript.
BF is supported by DOE grant DE-FG02-01ER-40676. ALF is supported by 
DOE grant DE-FG02-01ER-40676 and NSF CAREER grant PHY-0645456.

\appendix

\section{Dominance of the Neutrino Signal}

In this appendix, we will demonstrate that the neutrino signal is indeed the most
important signal in the models we have considered.  The main constraint comes
from the requirement that the dark matter decays must have a sufficiently small
branching ratio to produce positrons.  Even though this branching ratio is
very small, the positron background that the signal must overcome is smaller than the neutrino background by
a factor of order 1000, and moreover, the slow diffusion of positrons through
the galaxy results in a greater number of them remaining around to be
detected.  The dominance of the neutrino signal is thus something which must
be checked.  By making use of the ``leaky box'' model for cosmic ray
propagation, in which positrons are considered to undergo a random walk process due to the galactic magnetic field, we
will make a rough estimate for the required constraint on the
positron branching fraction.   This constraint will be seen to be satisfied by many
orders of magnitude in the models we consider, and for this reason, a more
precise calculation will not be necessary.

Let $Q_{e^+}(E)$ be the production rate per unit volume per unit energy for
positrons due to dark matter decays in the Milky Way.  Depletion of cosmic
ray positrons of energy $E \sim O($1-10 GeV$)$ occurs due to a variety of mechanisms; they may lose
energy- due to synchrotron radiation, inverse Compton scattering
 and bremsstrahlung- or they may actually manage to diffuse out of the galaxy.  An
equilibrium is established whereby the production and depletion effects
reach a balance, yielding the relation \cite{Kobayashi}
\begin{equation}
Q_{e^+}(E) \approx \frac{d}{dE}\left( \frac{dE}{dt} n_{e^+}(E)\right) + \frac{n_{e^+}(E)}{T}. 
\label{eq:Q}
\end{equation}
Here $n_{e^+}(E)$ is the number density per unit energy of signal positrons, and $T$
is the average time needed for a positron to diffuse out of the galaxy.
Energy losses due to sychrotron radiation and inverse Compton scattering
have $\frac{dE}{dt} \propto - E^2$.
 Bremsstrahlung, which begins to dominate at energies below a
few GeV, has  $\frac{dE}{dt} \propto - E$.  The net energy loss rate may then
be written as  
\begin{equation}
\frac{dE}{dt} = - b E^2 - c E,
\label{eq:dEdt}
\end{equation}
with $b \sim \frac{1}{2\times 10^5 \textrm{years} \times \textrm{TeV}}$ and $c
\sim \frac{1}{6 \times 10^7 \textrm{years}}$ \cite{Kobayashi, Stawarz}.  It follows then that positrons with energies of order 
several GeV will have a lifetime for energy loss of order $10^7$ years. 

The random walk process for the diffusing
positrons has them moving a typical distance
\begin{equation}
r = \sqrt{D \times t}
\label{eq:r}
\end{equation}
in a time t, where $D \sim 3 \times 10^{28}\textrm{cm}^2/\textrm{s}$.  The timescale
for a positron to diffuse out of the galaxy, and thus move a distance of
order the galactic height $10^3$ light years, is then given by 
\begin{equation}
T \sim  10^6 \textrm{years}.
\label{eq:T}
\end{equation}

We thus see that, at energies of order several GeV, 
the dominant depletion term on
the right hand side of equation (\ref{eq:Q}) is that due to 
diffusion.  
It then follows that $n_{e^+}(E)$ is given by
\begin{equation}
n_{e^+}(E) \sim Q_{e^+}(E) T
\label{eq:nE1}
\end{equation}

Let us assume for simplicity that the source $Q(E)$ is a delta function at an energy
$E_0$.  In reality decays involving positrons will involve many body final
states; this will only serve to spread out the spectrum and dilute the signal
somewhat.  We thus take $Q_{e^+}(E) = \frac{\rho_{DM}}{m_{DM}}
\Gamma_{DM} \gamma_{e^+} \delta(E - E_0)$, where $\gamma_{e^+}$ is the branching
ratio for dark matter decays to positrons.  Putting everything together, we obtain
\begin{equation}
n_{e^+}(E) \sim \gamma_{e^+} \frac{\rho_{\rm DM} \Gamma_{\rm DM}}
{m_{\rm DM}} T \delta(E-E_0)
\label{eq:nE2}
\end{equation}
For comparison, the dominant decays to neutrinos result in a spectrum (again
assuming a delta function source for simplicity, centered on $E_0$)
\begin{equation}
n_{\nu}(E) \sim \frac{\rho_{DM}\Gamma_{DM}}{m_{DM}} R_G \delta(E - E_0 ),
\label{eq:nNu}
\end{equation}
where $R_G$ is roughly the galactic scale of order $10^5$ light years.
Integrating over energy bins of size of order $E_0$,
 the ratio of positron to neutrino signals is therefore approximately
\begin{equation}
\frac{\int n_{e^+}(E)}{\int n_{\nu}(E)} \sim \gamma_{e^+} \frac{T}
{R_G} \sim 10 \gamma_{e^+}.
\label{eq:nratio}
\end{equation}
In our models we estimate the positron branching ratio (due to decays through
an off-shell W-boson) to be approximately $ \gamma_{e^+} \sim \frac{1}{(2
  \pi)^4}\frac{m_{DM}^6}{m_W^4 v^2}$, which is $ \sim 5 \times 10^{-10}$
at $m_{\rm DM} = 10$ GeV.  Given equation
(\ref{eq:nratio}), this easily allows
the neutrino signal to dominate, even given the $\sim 1000$ times larger
background flux in atmospheric neutrinos compared to cosmic ray positrons.

\label{app:dom}

\section{Annihilations}
\label{app:ann}

The framework for ADM we consider here cannot lead to a neutrino signal from 
dark matter annihilations in the galaxy at an observable level in the foreseeable 
future.  The reason is that a sufficiently large coupling of dark
matter to neutrinos would cause the lepton-asymmetry-transfering interactions
to freeze out below the dark matter mass.  To see this, recall that
the condition of freezing out above the DM mass can be written
\be
\sigma_m m_{\rm DM}^3  &\lesssim& H(m_{\rm DM}),
\ee
where $\sigma_T T^3 \sim \langle n \sigma v \rangle_T$ is the annihilation rate
at temperature $T$.
We therefore must have
\be
\frac{\Gamma_0}{m_{\rm DM}} &=& \frac{\sigma_0}{\sigma_m}
\sigma_m  \frac{\rho_{\rm DM}}{m_{\rm DM}^2}
  < \frac{\sigma_0}{\sigma_m} \sqrt{g_*(m_{\rm DM}) }
    10^{-62}
 \left( \frac{1\textrm{GeV}}
{m_{\rm DM}}\right)^3,
\ee
%
where $\Gamma_0$, $\sigma_0$ are the annihilation rate and
cross-section $\langle \sigma v \rangle $ today and $g_*(T)$ is
the effective number of relativistic degrees of freedom in the universe
at temperature $T$. One can see 
from Figure \ref{fig:decaybound} that absurd enhancements of the cross-section
over the value at momenta near the DM mass would be required
in order to give an observable rate of DM annihilations in the galaxy.

\end{document}